\documentclass[12pt]{article}
\usepackage{epsfig}

\def\Title#1{\begin{center} {\Large {\bf #1} } \end{center}}

\newcommand\beq{\begin{equation}}
\newcommand\eeq{\end{equation}}
\newcommand\beqa{\begin{eqnarray}}
\newcommand\eeqa{\end{eqnarray}}

\newcommand\bgamma{\mbox{\boldmath$\gamma$}}

\newcommand\bSigma{\mbox{\boldmath$\Sigma$}}

\newcommand\bk{{\bf k}}
\newcommand\bq{{\bf q}}
\newcommand\E{\epsilon}

\newcommand\bp{{\bf p}}

\begin{document}

\Title{Ferromagnetic properties of quark matter\\
- an origin of magnetic field in compact stars -}

\bigskip\bigskip


\begin{raggedright}

{\it Toshitaka Tatsumi\index{Tatsumi, T.}\\
Department of Physics\\
Kyoto University\\
Kyoto 606-8502\\
Japan\\
{\tt Email: tatsumi@ruby.scphys.kyoto-u.ac.jp}}
\bigskip\bigskip
\end{raggedright}

\section{Introduction}

The phase diagram of QCD in the density $(\rho_B)$ - temperature $(T)$ plane 
has been explored by many authors; QGP in high-$T$ region or color superconductivity in high-$\rho_B$ region is a typical phase in that plane. Here we are interested in low temperature and moderate density region relevant to compact stars, where magnetic order is expected.

Origin of the magnetic field in compact stars is one of the long-standing 
problems since the first discovery of pulsars in early seventies. 
Recent discovery of magnetars with huge magnetic field of $O(10^{14-15}{\rm G})$has revived this problem. Since nuclear matter is developed inside compact stars, 
we are tempted to consider spontaneous spin polarization of nucleons as a microscopic origin of the magnetic field. Realistic calculations have been performd for polarized nuclear matter, but they lead us to the negative results \cite{bom}. In ref.\cite{tat00} possibility of spin polarization of quark matter has been suggested by a simple consideration in analogy with itinerant electrons, where the Fock exchange interaction is responsible to ferromagnetism \cite{her}. A weakly first-order phase transition has been demonstrated around nuclear density. Using this result we can roughly estimate the magnitude of the magnetic field at the surface to be $O(10^{15-17}{\rm G})$, which may explain the magnetic field of magnetars. The coexistence of ferromagnetism with color superconductivity has been also discussed in ref.~\cite{nak03}.

Here we apply the Landau Fermi-liquid theory (FLT) to elucidate the
critical behavior of the magnetic phase transition at finite density
and temperature \cite{bay}. We evaluate the magnetic susceptibility of quark matter. The divergence and sign change of the magnetic
susceptibility is a signal of the magnetic instability to the ferromagnetic
phase, since its inverse measures the curvature of the free energy at
the origin with respect to the magnetization. 
Thus quarks near the Fermi surface are
responsible to 
the magnetic transition and the spin dependent quark-quark interaction and the density of states near the Fermi
surface are the key ingredients within FLT.

Theoretically we find a non-Fermi-liquid behavior of the magnetic susceptibility.It is well known that there appears a non-Fermi-liquid behavior in the expression of the specific heat in QCD as well as QED, which is caused by the transverse gauge field because it is not statically screened \cite{ipp}.

\section{Relativistic Fermi liquid theory}

Within the Landau Fermi-liquid theory (FLT) we assume a one-to-one correspondence
between the states of the free Fermi gas and those of the interacting
system \cite{bay}. Quarks are treated as quasi-particles carrying the same quantum
numbers of the free quarks, and the quasi-particle distribution function is simply given by the Fermi-Dirac one,
\beq
n(\bk,\zeta)=\left[1+\exp(\beta(\epsilon_{\bk,\zeta}-\mu))\right]^{-1}
\eeq
with the quasi-particle energy $\epsilon_{\bk,\zeta}$ specified by the
momentum $\bk$ and a spin quantum number $\zeta=\pm 1$.

\subsection{Screening effect on the quasi-particle interaction}

In the following we consider the color-symmetric interaction among
quasi-particles that can be written as the sum of two parts, 
the spin independent ($f^s_{\bk,\bq}$) and
dependent ($f^a_{\bk,\bq}$) terms;
\beq
f_{\bk\zeta,\bq\zeta'}=f^s_{\bk,\bq}+\zeta\zeta'f^a_{\bk,\bq}.
\eeq
Since quark matter is color singlet as a whole, the Fock exchange
interaction gives a leading contribution. We, hereafter,
consider the one-gluon-exchange interaction (OGE). For a pair with color
index $(a,b)$, the Fock exchange interaction gives a factor 
$(\lambda_\alpha)_{ab}(\lambda_\alpha)_{ba}=1/2-1/(2N_c)\delta_{ab}$, which 
is always positive for any pair. Hence the situation is very similar to
electron gas in QED. Since we are interested in the electromagnetic
properties of quark matter, only the color symmetric interaction is 
relevant, which is written as 
\beq
f_{\bk\zeta,\bq\zeta'}
=\frac{1}{N_c^2}\sum_{a,b}f_{\bk\zeta
a ,\bq\zeta' b }
=\frac{m}{E_k}\frac{m}{E_q}M_{\bk\zeta,\bq\zeta'},
\eeq
with the invariant matrix element,
\beq
M_{\bk\zeta,\bq\zeta'}=-g^2\frac{1}{N_c^2}{\rm
tr}\left(\lambda_\alpha/2\lambda_\alpha/2\right)M^{\mu\nu}(k,\zeta; q,\zeta')D_{\mu\nu}(k-q),
\eeq
where $M^{\mu\nu}(k,\zeta; q,\zeta')
={\rm
tr}\left[\gamma^\mu\rho(k,\zeta)\gamma^\nu\rho(q,\zeta')\right]$. 

Since the OGE interaction is a long-range force and we
consider the small energy-momentum transfer between quasi-particles, we
must treat the gluon propagator by taking into account HDL 
resummation. Thus we take into account the screening effect,
\beq
D_{\mu\nu}(k-q)=P^t_{\mu\nu}D_t(p)+P^l_{\mu\nu}D_l(p)-\xi\frac{p_\mu
p_\nu}{p^4} 
\eeq
with $p=k-q$, where $D_{t(l)}(p)=(p^2-\Pi_{t(l)})^{-1}$, and 
the last term represents the gauge dependence with a parameter
$\xi$. $P^{t(l)}_{\mu\nu}$ is the projection operator onto the
transverse (longitudinal) mode, 
\beqa
P^t_{\mu\nu}&=&(1-g_{\mu 0})(1-g_{\nu
0})\left(-g_{\mu\nu}-\frac{p_\mu p_\nu}{|\bp|^2}\right)\nonumber\\
P^l_{\mu\nu}&=&-g_{\mu\nu}+\frac{p_\mu
p_\nu}{p^2}-P^t_{\mu\nu}.
\eeqa
The self-energies for the transverse and longitudinal gluons are given
as 
\beqa
\Pi_l(p_0,{\bf p})&=&\sum_{f=u,d,s} \left( m_{D,f}^2+i\frac{\pi
m_{D,f}^2}{2u_{F,f}}\frac{p_0}{|\bp|} \right)\nonumber\\
\Pi_t(p_0,{\bf p})&=&-i\sum_{f=u,d,s} \frac{\pi u_{F,f} m_{D,f}^2}{4}\frac{p_0}{|\bp|}, 
\eeqa
in the limit $p_0/|\bp|\rightarrow 0$, with $u_{F,f}\equiv
k_{F,f}/E_{F,f}$ and the Debye mass for each flavor,
$m_{D,f}^2\equiv g^2\mu_f k_{F,f}/2\pi^2$ 
. 
Thus the longitudinal gluons are statically
screened to have the Debye mass, while the transverse gluons are dynamically screened by the Landau damping,
in the limit $p_0/|{\bf p}|\rightarrow 0$. Accordingly, the screening effect for the transverse gluons is ineffective 
at $T=0$, where soft gluons ($p_0/|\bp|\rightarrow 0$) contribute. 
At finite temperature, gluons with $p_0 \sim O(T)$ can contribute due to the diffuseness of the Fermi surface and the transverse gluons
are effectively screened.


\subsection{Magnetic susceptibility}

We consider the linear response of the normal(unpolarized) quark matter
by applying a small magnetic field 
$\bf B$. Using the Gordon identity, the coupling term with the uniform magnetic field (${\bf A}={\bf B}\times {\bf r}/2$) can be written as
\begin{eqnarray}
\int d^4x {\cal L}_{\rm int}&=&e_q\int d^4x{\bar q}\bgamma\cdot {\bf A}q\nonumber\\
&=&\mu_q\int d^4x {\bar q}\left[{\bf L}+\bSigma\right]\cdot{\bf B}q,
\end{eqnarray}
with $\mu_q$ being the Dirac magneton. We discard the contribution of the orbital angular momentum $\langle\bf L\rangle$ by assuming the uniform distribution of quarks. Thus the magnetization $\langle M\rangle_f$ for each flavor can be written as 
\beq
\langle M\rangle_f=V^{-1}\langle\int d^3x{\bar q}_f\Sigma_z q_f\rangle,
\eeq
where we take ${\bf B}//\hat {\bf z}$.
Accordingly the magnetic susceptibility is defined as
\beq
\chi_M=\sum_{f=u,d,s}\chi_M^f=\sum_{f=u,d,s}
\left.\frac{\partial\langle M\rangle_f}{\partial B}\right|_{B=0}.
\eeq
Hereafter, we shall concentrate on one flavor and omit the flavor indices 
because the magnetic susceptibility is given by the sum of the contribution from each flavors. The magnetic susceptibility is proportional to the number difference between different spin states ($\zeta=\pm 1$); it is explicitly caused by the applying magnetic field, and implicitly caused through the spin-dependent interaction,
\beq
\delta n_{\bk\zeta=+1}-\delta n_{\bk \zeta=-1}=\frac{\partial n_\bk}{\partial\epsilon_\bk}\left[-g_D\mu_qB+\delta\epsilon_{\bk\zeta=+1}-\delta\epsilon_{\bk \zeta=-1}\right]
\eeq
 with the gyromagnetic ratio $g_D\sim 2$, where 
\beq
\delta\epsilon_{\bk\zeta}=N_c\sum_{\zeta'=\pm 1}\int\frac{d^3 q}{(2\pi)^3}f_{\bk\zeta;\bq\zeta'}\delta n_{\bq\zeta'}.
\eeq
 Magnetic susceptibility is then written in terms of the quasi-particle 
interaction,
\beqa
\chi_M=\left(\frac{\bar g_D\mu_q}{2}\right)^2\frac{N(T)}{1+N(T)\bar
f^a}
\eeqa
 where $\bar g_D$ is an angle average of $g_D$, and $\bar f^a$ is the Landau-Migdal parameter averaged over the Fermi surface \cite{tat083,sat}.



\section{Magnetic properties at $T=0$}

$N(T)$ is the effective density of states at the Fermi surface,
and is simply written as 
\beq
N^{-1}(0)=\frac{\pi^2}{N_ck_F^2}v_F
\label{velocity}
\eeq
in the limit of zero temperature . Eq.~(\ref{velocity})
defines the Fermi velocity, which is given by using the Lorentz
transformation \cite{bay},
\beq
v_F\equiv \left.\frac{\partial
n_\bk}{\partial\epsilon_\bk}\right|_{|\bk|=k_F}
=\frac{k_F}{\mu}-\frac{N_ck_F^2}{3\pi^2}f_1^s,
\eeq 
 where $f_1^s$ is a spin-averaged Landau-Migdal parameter.

Finally the magnetic susceptibility at zero temperature can be written in terms of the
Landau-Migdal parameters,
\beq
\chi_M=\chi_{\rm Pauli}\left[1+\frac{N_ck_F\mu}{\pi^2}\left(-\frac{1}{3}f_1^s+\bar
f^a\right)\right]^{-1},
\eeq
where $\chi_{\rm Pauli}$ is the usual one for the Pauli paramagnetism,
$
\chi_{\rm Pauli}={\bar g}_D^{2}\mu_q^2N_ck_F\mu/4\pi^2.
$

The quasiparticle interaction on the Fermi surface can be written as 
\beqa
\left.f_{\bk\zeta,\bq\zeta'}\right|_{|\bk|=|\bq|=k_F}&=&-C_g\frac{m^2}{E_F^2}
\left[-M^{00}D_L(\bk-\bq)+M^{ii}D_T(\bk-\bq)\right],
\eeqa
with the effective coupling strength, 
$C_g=\frac{N_c^2-1}{2N_c^2}g^2$.

We can see that the both Landau parameters $f_1^s,
{\bar f}^a$ include
the infrared singularities due to the absence of the static screening
for the transverse gluons; $D_T(\bk-\bq)\sim
-1/(\bk-\bq)^2=-1/2k_F^2(1-\cos\theta_{\hat{\bk\bq}})$ in this case, so
that the logarithmic divergences appear in the Landau parameters through
the integral over the relative angle, $\int
d\Omega_{\hat{\bk\bq}}1/(1-\cos\theta_{\hat{\bk\bq}})$.

Finally magnetic susceptibility is given as
a sum of the contributions of the bare interaction and the static 
screening effect. We can see that the logarithmic divergences exactly 
cancel each other to give a finite result for susceptibility 
\cite{tat08,tat082}.
\begin{equation}
\left(\chi_M/ \chi_{\rm Pauli}\right)^{-1}=1
-\frac{C_gN_c\mu}{12\pi^2E_F^2k_F}\Big[m(2E_F+m)
-\frac{1}{2}(E_F^2+4E_Fm-2m^2)
\kappa\ln\frac{2}{\kappa} \Big],
\label{final}
\end{equation}
with $\kappa=m_D^2/2k_F^2$.
Obviously this expression is reduced to the simple OGE case without
screening in the limit $\kappa\rightarrow 0$; 
one can see that the interaction among massless quarks gives 
a null contribution for the magnetic transition. 
The effect of the static screening for the
longitudinal gluons gives the contribution of $g^4\ln(1/g^2)$. In the
nonrelativistic limit, it recovers the corresponding term in the RPA
calculation of electron gas \cite{her,tat08,tat082}.

\begin{figure}[htb]
\begin{center}
\epsfig{file=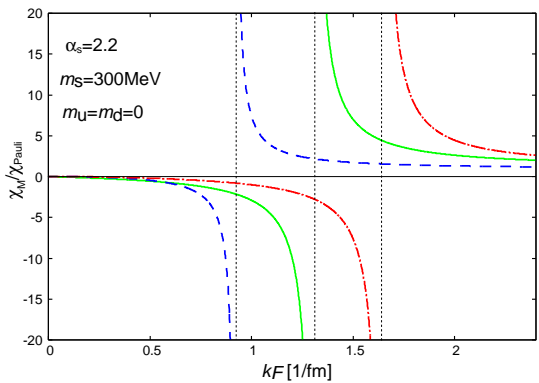,height=2in}
\caption{Magnetic susceptibility at $T=0$. The solid curve shows the
 result using simple OGE, 
while the dashed and dash-dotted ones show the screening effects with $N_f=1$(only s quarks) and with $N_f=3$(u, d, and s quarks) respectively.}
\label{fig:chi}
\end{center}
\end{figure}

In Fig. 1, we plot the magnetic susceptibility at $T=0$\cite{tat08,tat082}.
We 
take the QCD coupling constant as
$\alpha_s\equiv g^2/4\pi=2.2$ and the strange quark mass $m_s=300$MeV inferred from the MIT bag model. 
We consider here the MIT bag model as an effective model succeeded in
reproducing the low-lying hadron spectra. The coupling constant looks
rather large, but this value is required for the color magnetic
interaction to explain the mass splitting
of hadrons with different spins; e.g. for nucleon and $\Delta$
isobar. We think this feature is relevant in our study, because 
the coupling constant is closely related to the strength
of the spin-spin interaction between quarks in this model. 
Moreover, the quark density in the MIT bag model is moderate,  
$0.25$fm$^{-3}$, which is the similar one we are interested in.
Note that the perturbation method should be still meaningful even for this
rather large coupling, since the renormalization-group analysis has
shown that the relevant expansion parameter is not the gauge coupling
constant $g^2$ but the product of $g^2$ with the Fermi velocity $v_F$, 
which {\it always} goes to zero as one approaches to the Fermi surface \cite{sch}.  

One can see that the magnetic susceptibility for the simple OGE without screening
diverges around $k_F=1.3$ fm $^{-1}$.
This is consistent with the previous result for the energy calculation. \cite{tat08,tat082}.
One may expect that the screening effect weakens the Fock exchange interaction so that 
the critical density get lower once we take into account the screening effect. However, this is not
necessarily the case in QCD. The screening effect behaves in different ways depending on the number of flavors.
Compare the results for the $N_f=3$ with the one for $N_f=1$. In the case of $N_f= 1, \kappa \leq 2$ the screening effect 
works against the magnetic phase transition as in QED. 
However, for $N_f=3$, $\kappa>2$ so that the critical density is increased.
Consequently the screening effect does not 
necessarily work against the magnetic instability, which is a different
aspect from electron gas \cite{tat082}.

\section{Finite temperature effects and Non-Fermi-liquid behavior}

At finite temperature, the magnetic susceptibility is given by
\begin{equation}
\chi_M=\left(\frac{\bar g_D\mu _q}{2}\right)^2\left[N^{-1}(T)+\bar f_{ l}^a+ \bar f_{ t}^a\right]^{-1} \label{eq:5.31.5}
\end{equation}
where $\bar f_{ l}^a$ and $\bar f_{t}^a$ denote the longitudinal and transverse parts of $\bar f^a$ respectively \cite{tat083,sat}. First, we evaluate the effective density of states on the Fermi surface defined by
\beq
N(T)=\frac{N_c}{\pi^2} \int_{\E_0}^{\infty} d\omega  
\frac{d k}{d \omega} k^2 \frac{\beta e^{\beta\left(\omega-\mu \right)}}{\left(e^{\beta\left(\omega-\mu \right)}+1\right)^2}  , \label{eq:3.1.1}
\eeq
with $\E_0 \equiv \E_{|\bk|=0}$. The quasi-particle energy $\omega$
should be given as a solution of the equation,  
\beq
\omega = E_{k(\omega)} +{\rm Re}\Sigma_+(\omega,k(\omega)), \label{eq:omega}
\eeq 
where we discard the imaginary part within the quasi-particle
approximation.

The one-loop self-energy is almost independent of the momentum, and can be written
 as \cite{man}
\beq
{\rm Re}\Sigma_+(\omega,k)\sim
{\rm Re} \Sigma_+(\mu ,k_F)-\frac{C_fg^2u_{F}}{12\pi^2}(\omega-\mu )\ln\frac{\Lambda}{|\omega-\mu |} 
+\Delta^{\rm reg}(\omega-\mu)
\label{eq:Sigma}
\eeq 
around $\omega \sim \mu $ with $C_f=(N_c^2-1)/(2N_c)$ and $u_F=k_F/E_{k_F}$. $\Lambda$ is a
cut-off factor and should be an order of the Debye mass, $\Lambda\sim O(m_D)$.  Note that the
 anomalous term in Eq.~(\ref{eq:Sigma}) appears 
from the dynamic screening of the transverse gluons, and the
 contribution by the longitudinal gluons is summarized  
in the regular function $\Delta^{\rm reg}(\omega-\mu)$ of $O(g^2)$. 
Within the approximation given by Eqs. (\ref{eq:Sigma}) and (\ref{eq:ImSigma}), the self-energy is independent
of spatial momentum $k$  and thus we omit the argument $k$
 hereafter. The renormalization factor $z_+(k)$ is then given by the
 equation,
 $z_+(k)=(1-\partial{\rm Re}\Sigma_+(\omega)/\partial\omega|_{\omega=\epsilon_\bk})^{-1}$, and we have 
\beq
z_+(k)^{-1}\sim -\frac{C_fg^2u_F}{12\pi^2}\ln|\epsilon_\bk-\mu|.
\eeq
It exhibits a logarithmic divergence as $\epsilon_\bk\rightarrow \mu$,
 which causes non-Fermi liquid behavior \cite{sch}.

Eventually, $N(T)$ is written as,
\beq
N(T) \simeq \frac{N_c}{\pi^2} \int_{\E_0}^{\infty} d\omega \left(1-
 \frac{\partial{\rm Re} \Sigma_+(\omega)}{\partial
 \omega}\right)k(\omega)E_{k(\omega)} 
\frac{\beta e^{\beta\left(\omega-\mu \right)}}{\left(e^{\beta\left(\omega-\mu \right)}+1\right)^2} .\label{eq:NTomega}
\eeq
We can separate the contribution by the
 longitudinal gluons $N_l(T)$ from $N(T)$. Since the longitudinal gluon
 exchange is short-ranged by the Debye screening mass, it becomes
 almost temperature independent,   
\beq
N_l(T)=\frac{N_ck_FE_F}{3\pi^2}f_{l;1}^s,
\label{NTl}
\eeq
with the Landau-Migdal parameter $f_{l;1}^s$,
\begin{eqnarray}
f_{l;1}^s=-\frac{3N_c^{-1}C_fg^2}{8E_F^2k_F^2}\left[\kappa k_F^2+2E_F^2\right]
\left[(1+\kappa)I_0(\kappa)-1\right],
\label{f1ls}
\end{eqnarray}
where $\kappa=\sum_f m_{D,f}^2/2k_F^2$ and
\begin{equation}
I_0(\kappa)=\frac{1}{2}\int_{-1}^1\frac{du}{1-u+\kappa}
\simeq \frac{1}{2}{\rm
      ln}\left(\frac{2}{\kappa}\right)
\simeq {\rm ln}(g^{-2}).
\end{equation}

To evaluate the transverse contribution, $N_t(T)=N(T)-N_l(T)$, we only use
 the transverse part in Eq.~(\ref{eq:Sigma}):
substituting Eq. (\ref{eq:Sigma}) into Eq. (\ref{eq:NTomega}), we obtain 
the leading order contribution
\footnote{We discard here the temperature independent term of $O(g^2)$,
 which cannot be given only by Eq. (\ref{eq:Sigma}). However, we can
 recover it by taking the $T\rightarrow 0$ limit later. 
}
,
\begin{eqnarray}
N_t(T)&=&\frac{N_ck_s\mu }{\pi^2}\Big[
 1+\frac{\pi^2}{6}\frac{(2k_F^2-m^2)}{k_F^4}T^2 \nonumber\\
&&+\frac{C_fg^2u_F}{24}\frac{(2k_F^2-m^2)}{k_F^4}
T^2\ln\left(\frac{\Lambda}{T}\right)
 +\frac{C_fg^2u_F}{12\pi^2}\ln\left(\frac{\Lambda}{T}\right)
 \Big]+O(g^2T^2)
,\label{eq:NT1}
\end{eqnarray} 
after some manipulation.
$N_t(T)$ has a term proportional to $\ln T$, which gives a singularity at
 $T=0$. This singularity corresponds to 
the logarithmic divergence of the Landau-Migdal parameter $f_1^s$ at
 $T=0$. The chemical potential $\mu$ in Eq. (\ref{eq:NT1}) implicitly includes the
temperature dependence. To extract the proper temperature dependence in
$\chi_M$ we must carefully take into account the temperature dependence
of $\mu$. Using the thermodynamic relation $\mu=-(\partial
F/\partial n)|_T$ with the free energy $F=E-Ts$, we have \cite{sat} 
\begin{equation}
\mu(T)=\mu_0-\frac{\pi^2}{6}\frac{2k_F^2+m^2}{k_F^2E_F}T^2\left(1+\frac{C_fg^2u_F}{12\pi^2}
\ln\left(\frac{\Lambda}{T}\right)\right)+O(g^2T^2).
\label{mut}
\end{equation}
We can see that $\mu$ 
includes $T^2\ln T$ term due to the dynamic screening effect for the transverse gluons, besides the usual $T^2$ term. 

As for the spin-dependent Landau-Migdal parameter, the leading-order contribution at finite temperature comes from the
transverse component $\bar f_{ t}^a$; it has a logarithmic
singularity at $T=0$ due to the dynamic screening effect.
In this section, we shall see that 
the logarithmic divergences of
$N^{-1}(T)$ and $\bar f_{ t}^a$ at $T=0$ cancel out each other to
give a finite contribution to the magnetic susceptibility.
$\bar f_{ t}^a$ is given by
\beq
 \bar f_{ t}^a=-2N_c N^{-1}(T)\int\frac{d^3k}{(2\pi)^3}\frac{\partial n(\epsilon_\bk)}{\partial \epsilon_{\bk}} \bar f^a_{{t};k,k_s} \label{eq:5.32}
\eeq
with
\beq
\bar f^a_{t:k,k_s} =
 -\left.\int\frac{d\Omega_\bk}{4\pi}\int\frac{d\Omega_\bq}{4\pi}
 \frac{m^2}{E_s E_\bk}C_fN_c^{-1}g^2M^{iia} D_{ t}(k-q)\right|_{|\bq|=k_s}  \label{eq:5.35}
\eeq
where $M^{iia}$ is the spin-dependent component of $M^{ii}$ in 
Eq.(4), and $k_s=k_F+O(T^2)$ is defined by $\epsilon_{k_s}=\mu$.

The real part of the transverse propagator is
\beq
{\rm Re} D_{t}(k-q)\Big
 |_{|\bq|=k_s}=\frac{(k-q)^2}{\left\{(k-q)^2\right\}^2+\left(\frac{1}{4}\sum_f \pi u_{F,f} 
m_{D,f}^2\right)^2\frac{(E_\bk-E_s)^2}{(\bk-\bq)^2}}\Bigg|_{|\bq|=k_s} 
\eeq 
, while the imaginary part gives only a sub-leading contribution and can be discarded.

The integral over $k$ in Eq. (\ref{eq:5.32}) can be performed as in Eq. (21). 
Finally we find a leading-order contribution at $T\neq 0$, 
\begin{eqnarray}
\bar f_{t}^a &\sim& N^{-1}(T)\frac{C_fg^2}{12\pi^2k_sE_s}\left[
1 +\frac{\pi^2}{6}\frac{(2k_s^2-m^2)}{k_s^4}T^2 \right]\ln T^{-1} +O(g^2T^2) \nonumber \\ 
&\sim& \frac{C_fg^2}{12N_cE_s\mu}\ln T^{-1}. \label{eq:fas}
\end{eqnarray} 
Compare Eq. (\ref{eq:fas}) with Eq. (27). Since
$E_s=E_F+O(T^2)$ and $k_s=k_F+O(T^2)$ as we shall see, the $\ln T$ terms cancel each other in the magnetic
susceptibility (\ref{eq:5.31.5}).

\begin{eqnarray}
\left(\chi_M/\chi_{\rm Pauli}\right)^{-1} = 
1 -\frac{C_fg^2}{12\pi^2E_Fk_F}\Big[m(2E_F+m)%
-\frac{1}{2}(E_F^2+4E_Fm-2m^2)
\kappa\ln\frac{2}{\kappa} \Big] \nonumber\\
+\frac{\pi^2}{6k_F^4} \left(2E_F^2-m^2+\frac{m^4}{E_F^2} \right)T^2
+\frac{C_fg^2u_F}{72}\frac{(2k_F^4+k_F^2m^2+m^4)}{k_F^4E_F^2} T^2\ln\left(\frac{\Lambda}{T}\right) \nonumber\\
+O(g^2T^2). 
\label{chi_finT}
\end{eqnarray}

In Fig.2, we plot the magnetic susceptibility given by Eq. (\ref{chi_finT}).
At $T$=0, the magnetic susceptibility is positive at higher densities
and the quark matter is in 
the paramagnetic phase there. At the critical density where the magnetic
susceptibility diverges($k_F^c\sim 1.6$fm$^{-1}$), 
there occurs a magnetic phase transition from the paramagnetic phase to the ferromagnetic phase and the quark matter remains in the ferromagnetic phase 
below $k_F^c$.

At $T=$30 MeV, there appear two critical densities at which the magnetic
susceptibility  diverges. We denote these 
densities $k_{F1}^c$ and $k_{F2}^c$ ($k_{F1}^c<k_{F2}^c$). In this case,
$k_{F1}^c \simeq 0.4$fm$^{-1}$ and $k_{F2}^c \simeq 1.5$ fm$^{-1}$. 
At densities below $k_{F1}^c$ and above $k_{F2}^c$, the magnetic
susceptibility is positive, which corresponds to the paramagnetic phase, 
on the other hand, at densities between two critical densities, it becomes  negative corresponding to the ferromagnetic phase.

At $T=$50 MeV, there are still two critical densities ($k_{F1}^c\simeq
0.7$fm$^{-1}$ and $k_{F2}^c \simeq 1.3$fm$^{-1}$), 
but the range between these two densities becomes narrower than at $T=$30 MeV.

At $T=$60 MeV, there is no longer divergence in the magnetic susceptibility and quark matter is in the paramagnetic phase at any density.

\begin{figure}[htb]
\begin{center}
\epsfig{file=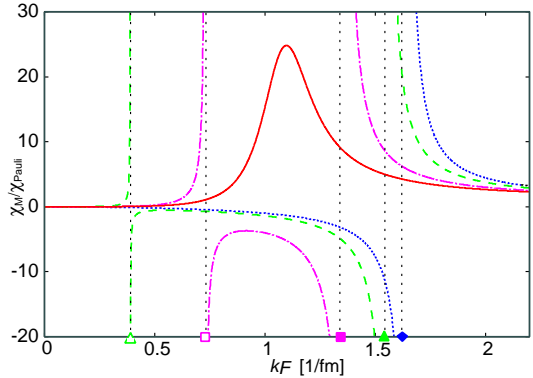,height=2in}
\caption{Magnetic susceptibility at finite temperature. 
The dotted, dashed, dash-dotted, and solid curves show the results at $T$=0, 30, 50, and 60 MeV respectively.
}
\label{fig:temp}
\end{center}
\end{figure}

We show a magnetic phase diagram of QCD on the density-temperature plane in Fig.3.
The four curves corresponds to the critical curves given by
Eq.(\ref{chi_finT}) 
under four different assumptions: below the curves the quark matter is
in the ferromagnetic phase, 
while it is in the paramagnetic phase above the critical curves. The magnetic transition occurs on the critical curves.

For the solid curve, we have used the full expression
Eq.(\ref{chi_finT}), on the other hand, 
for the dashed, dash-dotted, and dotted curves, we have ignored the
dynamic screening({\it i.e.} the $T^2 \ln T$ term), 
static screening({\it i.e.} the $\kappa \ln \kappa$ term), and both of the two screenings in Eq. (\ref{chi_finT}) respectively.
 
Compare the result with the full expression (\ref{chi_finT}) with the
one without 
the non-Fermi-liquid effect {\it i.e.} $T^2 \ln T$ dependence. In the
case without the $T^2 \ln T$ term, 
the ferromagnetic phase can be sustained till over $T=60$ MeV, while it can be at most $T=60$MeV including $T^2 \ln T$ dependence.
It turns out that the dynamic screening works against the magnetic
instability and can reduce the ferromagnetic region in the phase diagram up to a point,
but this effect is not so large.

The dash-dotted curve is the result without the static screening or
$\kappa \ln \kappa$ term in Eq.(\ref{chi_finT}). The static screening
effect works in favor of the magnetic instability to enlarge the
ferromagnetic region. As discussed in ~\cite{tat082}, it depends on the
number of flavors whether the static screening works for the 
ferromagnetism or not, which is peculiar to QCD.

\begin{figure}[htb]
\begin{center}
\epsfig{file=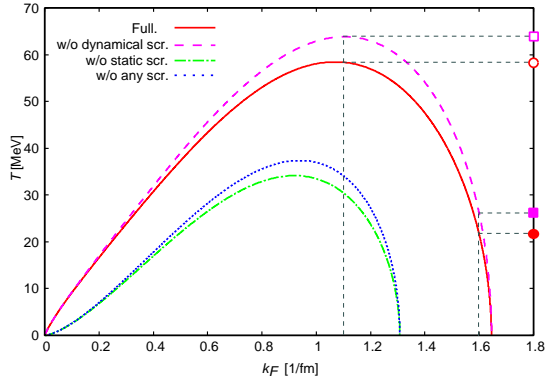,height=2in}
\caption{Magnetic phase diagram in the density-temperature plane. The
 solid, dashed, dash-dotted, dotted curves show the results for the full
 expression Eq. (\ref{chi_finT}), the one without the $T^2 \ln T$ term,
 without the $\kappa \ln \kappa$ term, and without the $T^2 \ln T$ and
 $\kappa \ln \kappa$ terms in Eq. (\ref{chi_finT}). The open (filled)
 circle indicates 
the Curie temperature at $k_F=1.1(1.6)$ fm$^{-1}$ while the squares show those when we disregard the $T^2 \ln T$ dependence.}
\label{fig:phased}
\end{center}
\end{figure}

The maximum Curie temperature $T_c^{\rm max}$ is around $60$MeV, which is achieved at $k_F\simeq
1.1$fm$^{-1}$. Note that this is still low temperature, since $T_c^{\rm
max}/k_F\ll 1$. Thus our low-temperature expansion is legitimate over all
points on the critical curve. One of the interesting phenomenological
implications may be related to thermal evolution of magnetars;
during the supernova expansions
temperature rises up to several tens MeV, which is so that ferromagnetic phase
transition may occur in the initial cooling stage to produce huge
magnetic field.

\section{Outlook}

We have discussed the critical behavior of the magnetic susceptibility in the density-temperature plane within the Fermi liquid theory. We have found a novel non-Fermi-liquid behavior and phase boundary by a perturbative calculations.
Some non-pertubative effects such as instanton effects should be taken into account at moderate densities. This is important not theoretically but also phenomenologically; more realistic estimate of the critical density or the Curie temperature is needed when we face phenomena in compact stars.

There are various ideas such as amplification of the fossil field for the origin of the magnetic field in compact stars. So 
it should be very interesting if we can distinguish these ideas through observations. To this end we must consider not only magnetic evolution but also thermal evolution; if ferromagnetic state is realized, spin waves should be excited which affect the thermal evolution of compact stars \cite{tat08}.

\bigskip
This work was partially supported by the 
Grant-in-Aid for the Global COE Program 
``The Next Generation of Physics, Spun from Universality and Emergence''
from the Ministry of Education, Culture, Sports, Science and Technology
(MEXT) of Japan  
and the 
Grant-in-Aid for Scientific Research (C) (20540267).

\def\Discussion{
\setlength{\parskip}{0.3cm}\setlength{\parindent}{0.0cm}
     \bigskip\bigskip      {\Large {\bf Discussion}} \bigskip}
\def\speaker#1{{\bf #1:}\ }
\def\endDiscussion{}





\end{document}